\documentstyle[aps,prl,floats,twocolumn,epsf]{revtex}

\newcommand{\LyX}{L\kern-.1667em\lower.25em\hbox{Y}\kern-.125emX\spacefactor1000}

\begin{document}
\noindent
{\bf Comment on ``Activation Gaps and Mass Enhancement of Composite Fermions''}

In a recent Letter\cite{pandj}, Park and Jain presented results of extensive
Monte Carlo calculations of gaps of polarized fractional quantum Hall states
at filling fractions \( \nu =1/3,2/5,3/7,4/9 \). In particular, they addressed
the effects on these excitation energies of the finite width of the electronic
wave function (wf) in the direction perpendicular to the interface. Their
findings appeared to demonstrate that finite width corrections alone lead to
agreement between experimental and theoretical values for these energy gaps.

This agreement depends crucially on the choice of the width parameter
\( \lambda  \) in the model interaction 
\begin{equation}
\label{eq1}
V(r)=\frac{e^{2}}{\epsilon \sqrt{r^{2}+\lambda ^{2}}},
\end{equation}
that was used to take account of the width of the interface wf
in an approximate manner. To fix \( \lambda  \), Park and Jain used results
based on calculations of interface wf's by Ortalano et al.\cite{Orta},
by fitting the gap at \( \nu =1/3 \) for a system of 6 electrons \cite{Orta}.
This fit leads to a value \( \lambda =11.7nm \)
for an electronic density \( n_{S}=2.3\times 10^{11}cm^{-2} \) of the sample
used by Du et al.\cite{Du}.
This value for \( \lambda  \) leads to the above mentioned agreement between
theoretical and experimental gap values
(cf. Figure 3 in \cite{pandj}).

The main problem with this way of fixing \( \lambda  \) is that there are indications
that the work of Ortalano et al. is not reliable\cite{morf_sarma}:
Analysis of the tabulated values of the Haldane pseudopotentials reveals serious
inconsistencies and the calculation of energy gaps is not explained in enough
detail to allow a verification of the results. Furthermore, as will be discussed
below, the value of \( \lambda  \) which fits the gap results does not appear
reasonable, although Park and Jain claim that it is.

To see this we employ as interface wf the Fang-Howard variational
wf \( \Psi (z)\sim \exp (-bz/2) \)\cite{ando}. The variational
parameter \( b \) is given by 
\begin{equation}
\label{eq2}
b=\left( \frac{48\pi m^{*}e^{2}(n_{depl}+\frac{11}{32}n_{S})}{\epsilon \hbar ^{2}}\right) ^{1/3},
\end{equation}
where \( m^{*} \) is the effective mass in the direction normal to the interface
and \( n_{depl} \) is the charge density in the depletion layer. Using the
values for GaAs, a dielectric constant \( \epsilon =12.7,
m^{*}=0.067m_{0} \) and
neglecting \( n_{depl} \), we obtain \( 1/b=3.1nm \) at \( n_{S}=2.3\times 10^{11}cm^{-2}. \)
As gaps at filling factors \( 1/3\leq \nu \leq 2/3, \) are basically
determined by the Haldane pseudopotential \( V_{m} \)\cite{Orta} for angular
momentum \( m=1 \), the best way to fix \( \lambda  \) is by requiring that
$V_1$ is reproduced when calculated with the model
interaction (\ref{eq1}). Using the Fang-Howard wf with $1/b=3.1nm$,
we obtain \( V_{1}=0.3712 e^2/(\epsilon \ell_0) \) at \( \nu =1/3 \)
(i.e. at a magnetic field $B=28.5T$, and a magnetic length
$\ell_0=4.8nm$). This $V_1$-value  is reproduced using the model
interaction (\ref{eq1}) with \( \lambda /\ell _{0}=1.19 \),
i.e. \( \lambda =5.7nm \),
about half the value used by Park and Jain \cite{ndepl}. 

The question of wf width has been explored experimentally by Willett
et al. \cite{willett88}. By measuring the energy difference between
the first and second subband, they obtained a more accurate estimate
of the subband wf's with the result \( 1/b=3.9nm \). As that sample
had a smaller density \( n_{S}=1.65\times 10^{11}cm^{-2} \), and assuming
\( b \) scales as \( n_{S}^{1/3} \), we obtain a value \( 1/b=3.5nm \)
for the sample considered by Park and Jain, only about 10 percent larger than
the value based on equation (\ref{eq2}). Again requiring that $V_1$ be
reproduced by the model interaction, at $\nu=1/3$, we 
obtain \( \lambda /\ell _{0}=1.34 \), still very much smaller than the
value  \( \lambda /\ell _{0}=2.5 \) used by Park and Jain. On the basis
of these results, we feel that
the value of \( \lambda =11.7nm \) used by Park and Jain is too large by 
about a  factor of two. 

Clearly, for \( \lambda /\ell _{0} \approx 1.3 \) the gap
reduction due to the width of the interface wf is only about half of
what Park and Jain obtained. Thus, a significant difference between
theoretical and experimental gap values remains. However,  effects of disorder
and Landau level mixing\cite{yoshioka} are expected to further reduce the
theoretical gap results.

A last comment regards the use of the model interaction (\ref{eq1}).
While it is possible to fix $\lambda$ such that the pseudopotential
coefficient $V_m$ is reproduced exactly for $m=1$, those for $m>1$
when calculated with the model interaction, are generally too
large, especially for large $\lambda$ \cite{Orta,morf_sarma}. Consequently,
since incompressibility of polarized states in the range $1/3 \leq \nu \leq 2/3$
depends on the ratio $V_3/V_1$  not becoming too large, 
a phase transition to compressible
states may be predicted erroneously if the model interaction (\ref{eq1}) is
used (cf. Figure 3 in \cite{pandj}). Realistic interactions based on
appropriate interface wf's must be used for a reliable study of this
question \cite{morf_sarma}.
\vskip 3mm
\noindent
Rudolf H. Morf\\
Condensed Matter Theory,\\
Paul Scherrer Institute,\\
CH-5232 Villigen, Switzerland


\begin{thebibliography}{}
\bibitem{pandj}K. Park and J.K. Jain, Phys. Rev. Lett. {\bf 81}, 4200 (1998)
\bibitem{Orta}M.W. Ortalano, Song He, S. Das Sarma, 
Phys. Rev. {\bf B55}, 7702 (1997)
\bibitem{Du}R.R. Du et al., Phys. Rev. Lett. {\bf 70}, 2944 (1993)
\bibitem{morf_sarma}R. H. Morf and S. Das Sarma, to be published
\bibitem{ando}T. Ando, A.B. Fowler and F. Stern, 
Rev. Mod. Phys. {\bf 54}, 437 (1982)
\bibitem{ndepl}If a non-zero value for the depletion density \( n_{depl} \) is
used, \( 1/b \) and consequently $\lambda$ will be reduced somewhat. 
\bibitem{willett88}R.L. Willett, H.L. Stormer, D.C. Tsui, A.C. Gossard and J.H. English, Phys.
Rev. {\bf B37}, 8476 (1988)
\bibitem{yoshioka}D. Yoshioka, J. Phys. Soc. Jpn. {\bf 55}, 885 (1986)
\end{thebibliography}
\end{document}